# Oxygen reduction mechanisms in nanostructured $La_{0.8}Sr_{0.2}MnO_3$ cathodes for Solid Oxide Fuel Cells

**Joaquín Sacanell**[*,1,2], **Joaquín Hernández Sánchez**[1], **Adrián Ezequiel Rubio López**[1,2], **Hernán Martinelli**[1], **Jimena Siepe**[1], **Ana G. Leyva**[1,3], **Valeria Ferrari**[1,2], **Dilson Juan**[1,2], **Miguel Pruneda**[4], **Augusto Mejía Gómez**[1,2], **Diego G. Lamas**[1,2,3]

[1]*Departamento de Física de la Materia Condensada, Gerencia de Investigación y Aplicaciones, Centro Atómico Constituyentes, Comisión Nacional de Energía Atómica, Av. General Paz 1499 (1650) Buenos Aires, Argentina.*
[2]*CONICET, Argentina.*
[3]*Escuela de Ciencia y Tecnología, Universidad Nacional de General San Martín, Martín de Irigoyen 3100, Edificio Tornavía, Campus Miguelete, (1650) San Martín, Pcia. de Buenos Aires, Argentina.*
[4]*Catalan Institute of Nanoscience and Nanotechnology (ICN2), CSIC and The Barcelona Institute of Science and Technology, Campus Bellaterra, 08193 Barcelona, Spain.*

*Corresponding author:
Joaquín Sacanell
Departamento de Física de la Materia Condensada, Gerencia de Investigación y Aplicaciones, Centro Atómico Constituyentes, Comisión Nacional de Energía Atómica.
Avenida General Paz 1499, San Martín (1650) Buenos Aires, Argentina.
+541167727657
sacanell@tandar.cnea.gov.ar*

**Abstract**

In this work we outline the mechanisms contributing to the oxygen reduction reaction in nanostructured cathodes of $La_{0.8}Sr_{0.2}MnO_3$ (LSM) for Solid Oxide Fuel Cells (SOFC). These cathodes, developed from LSM nanostructured tubes, can be used at lower temperatures compared to microstructured ones, and this is a crucial fact to avoid the degradation of the fuel cell components. This reduction of the operating temperatures stems mainly from two factors: i) the appearance of significant oxide ion diffusion through the cathode material in which the nanostructure plays a key role and ii) an optimized gas phase diffusion of oxygen through the porous structure of the cathode, which becomes negligible. A detailed analysis of our Electrochemical Impedance Spectroscopy supported by first principles calculations point towards an improved overall cathodic performance driven by a fast transport of oxide ions through the cathode surface.

## Introduction

Fuel cells are one of the most promising devices for environmentally clean power generation by directly converting chemical energy into electricity. Among them, Solid Oxide Fuel Cells (SOFCs) [1] stand out mainly due to their high efficiency and the possibility of using different fuels such as hydrocarbons or hydrogen. However, the typically high operating temperature of SOFCs (~1000ºC) leads to the degradation of the materials of the cell. Hence, significant effort has been devoted to obtain novel materials and structures that could operate at lower temperatures, both for the electrodes [2-14] and the electrolyte [15-18], mostly in the so called "intermediate temperature" (IT) range (500 - 700ºC).

The cathodes can be either electronic conductors (EC), capable of catalysing the oxygen reduction reaction (ORR) or mixed electronic and ionic conductors (MIEC). In the former case, the ORR can only take place at the triple phase boundary (TPB), the region where gas, electrolyte and electrode meet. To allow the access of oxygen gas to the TPB regions, the electrode must be highly porous which is crucial in the case of EC [19-22] and highly beneficial for MIEC [6-10].

The performance of the cathode strongly depends on its microstructure, as it was shown for MIEC materials [3-14]. In that line, we have shown in previous works that the use of submicrometric tubes and rods constituted by assembled nanoparticles enhances both gas access and ionic conductivity of the cathode [9,10]. However, the use of nanostructured EC cathodes has not been widely explored so far.

Microcrystalline $La_{0.8}Sr_{0.2}MnO_3$ (LSM) is one of the most used cathodes for SOFCs [19-25] despite its negligible ionic conductivity [26-30]. In contrast, nanostructured LSM cathodes have not been studied yet, even though their interest has been recognized [31]. The main obstacle for this kind of study is the high temperatures needed for a good performance of the cathode, which conspire against the nanostructure stability (i.e., nanocrystals are expected to grow during operation at high temperatures). However, there are two important facts reported in the literature that clearly demonstrate that nanostructured LSM cathodes deserve great attention. On one hand, studies by the isotope exchange depth profiling technique using SIMS depth profiling on microcrystalline $LaMnO_3$ and LSM show that diffusion of oxygen is much faster along grain boundaries than through the bulk [32,33]. On the other hand, recent studies on LSM thin films and nanopowders have reached to the same conclusion and, in addition, a faster oxygen exchange kinetics has been observed [34,35].

In this work, we evaluate for the first time the performance of nanostructured LSM cathodes on yttria-stabilized zirconia (YSZ) electrolytes under different operation temperatures and oxygen partial pressures by electrochemical impedance spectroscopy (EIS) using symmetric [LSM/YSZ/LSM] cells and analyse the mechanisms involved in the ORR. We show that the use of nanostructured LSM enhances the cathodic performance in SOFCs in comparison with microstructured ones. This enhancement is mainly due to both an improved contribution of gas phase transport, making its contribution negligible, and the appearance of significant oxide ion diffusion through the cathode. These improvements place our nanostructured LSM cathodes as promising candidates for IT-SOFCs.

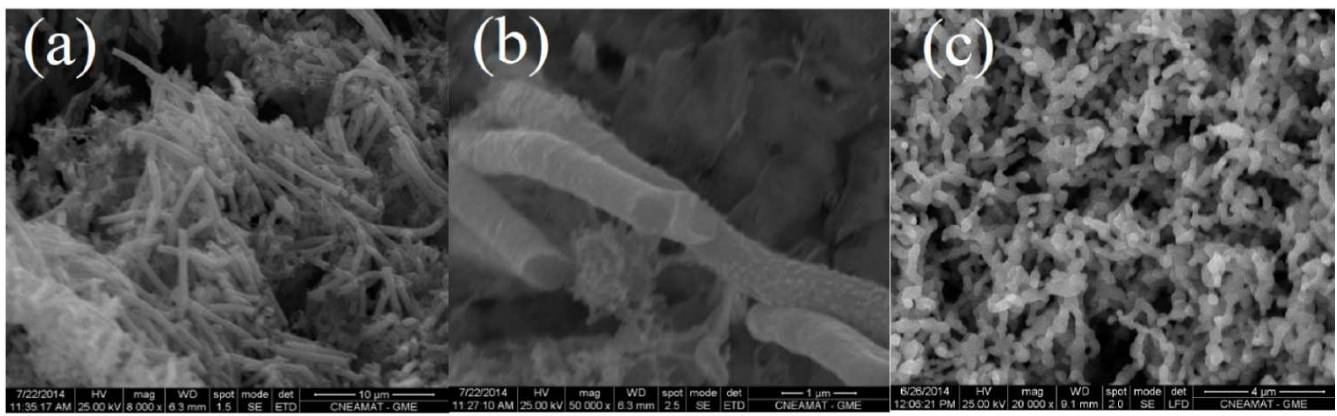

**Figure 1:** SEM images of: (a) Precursor LSM nanotubes, (b) a zoomed-in image of the precursors displaying the hollow structure. (c) Cathode after the sintering procedure.

**Experimental**

For LSM nanostructures, 1 M stoichiometric nitric solutions were prepared by dissolution of La, Sr and Mn nitrates in pure water. Porous polycarbonate films (Isopore$^{TM}$ from Millipore) of 400 nm pore diameter, were used as templates in a system for syringe filtration. The membranes were further treated under microwave radiation for a few minutes and then calcined at 900ºC for 10 min. The resulting powder is a collection of nanostructured tubes (Figures 1(a) and (b)) composed of an agglomeration of nanoparticles. A detailed description of the synthesis procedure and characterization can be found in reference [36]. The electrolyte substrates are pellets of $ZrO_2$-8%$Y_2O_3$ (YSZ) powder (Tosoh$^{TM}$) pressed into 10 mm diameter discs of about 1 mm thickness and sintered at 1600ºC for 4 hs.

The cathode deposition was performed by painting both sides of the electrolytes with an ink prepared with LSM nanotubes and a commercial ink vehicle (IV, Nextech Materials$^{TM}$), to obtain symmetrical cells. The sample was dried with an infrared lamp, and subsequently treated at 1000 ºC to attach the cathode. Samples with a cumulative number of depositions were made, in order to study the dependence of the properties with the thickness of the cathode.

Electrochemical Impedance Spectroscopy (EIS) measurements were performed as a function of temperature and the oxygen partial pressure, with a Gamry 750 potentiost-galvanostat-impedance and with Pt paste as current collector. Measurements were performed at zero bias with an amplitude of 20 mV. Polarization Resistance values were determined from EIS data.

**Calculations**

To complement the experimental results, we have performed ab initio calculations within Density Functional Theory (DFT) using the implementation of the SIESTA code [37]. Although the nanostructures could present different facets, we considered the (001) surface, which has the lowest energy [38]. For this surface, both the $MnO_2$ and LaSrO terminations are stable at room temperature [39] and even more, considering that nanostructuring should introduce disorder. However, we focus on $MnO_2$-terminated slabs, as Mn-terminated surfaces are more active for oxygen reduction compared to the (La,Sr)-terminated ones [40]. We considered a 2x2 periodicity at the surface and a 6x6x1 k-point mesh. The doping was included by explicitly substituting La atoms by Sr ones to reach the 0.2 composition. In all cases, the internal coordinates were allowed to relax up to forces smaller than 0.04 eV/Ang.

**Results and Discussion**

Fig. 1 (c) presents an image of the cathode, after the attachment procedure. The resulting cathode is formed by a collection of dense rods but the surface remains highly porous. It is worth noting that the obtained porosity is due to the morphology of the precursors and also to the fact that they are intertwined within the ink used for cathode deposition. As a consequence, grain growth is prevented due to the few contact points between the LSM particles, and thus the nanostructure is preserved.

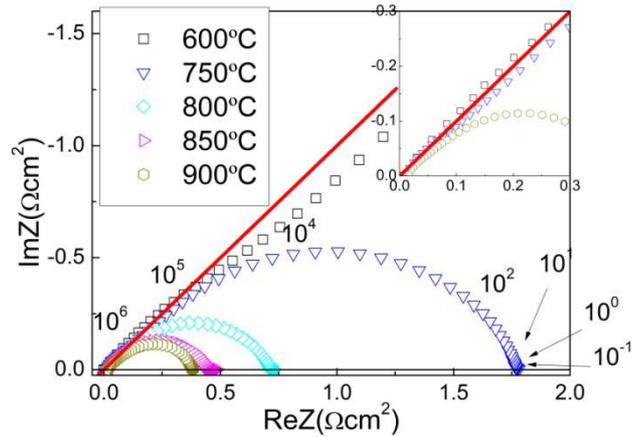

**Figure 2:** (color online) Nyquist plots for the nanostructured LSM cathode for different temperatures. The numbers close to the 750ºC data indicate the measurement frequency in Hz. The 45º slope solid line highlight the asymptotic behavior at high frequencies. Inset: detailed view of the high frequency part of the EIS spectra.

Fig. 2 shows the imaginary vs. the real part of the impedance Z (i.e. the Nyquist plot [41]) measured in air at different temperatures. With this technique, the different cathodic processes can be separated by their corresponding characteristic times. The first significant feature is the asymptotic 45º slope at high frequencies (shown as a line in Fig. 2), which is a signal of a contribution to the EIS spectra called Warburg element [41]. This element is generally attributed to solid state diffusive processes or ionic conduction and is typically observed in MIEC cathodes. For that reason, it has not been observed in pure electronic conductors such as LSM. The asymptotic behaviour clearly observed for temperatures below 750ºC, is progressively lost on heating, showing that the process leading to the Warburg element becomes less dominant when increasing the temperature as can be seen in the inset of Fig. 2.

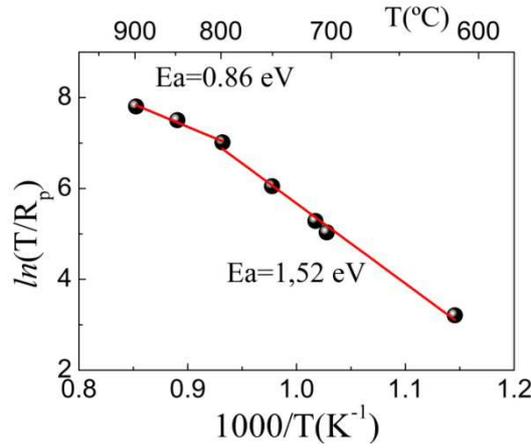

**Figure 3:** (color online) Arrhenius plot of $ln(T/R_p)$ ($R_p$: polarization resistance) and activation energies above and below 800ºC.

From the Nyquist plot, we can obtain the Polarization Resistance ($R_p$) of the cathodes, as the difference between the intersection of low and the high frequency data with the real axis. It is important to note that the $R_p$ values are smaller than 1 $\Omega.cm^2$ for temperatures above 750ºC. Those values are close to those obtained for microstructured cathodes with intrinsic ionic conduction, such as LSM-YSZ composites [20-24].

In a (T/$R_p$) graph vs. the inverse of temperature, the following Arrhenius relation can be assumed [1]:

$$\frac{T}{R_p} = \frac{1}{R_{p0}} exp\left(-\frac{E_a}{k_B T}\right) \qquad (1)$$

where $R_{p0}$ is a constant pre-exponential factor, $T$ is the absolute temperature, $k_B$ is the Boltzmann constant and $E_a$ is the activation energy of the corresponding process. In Fig. 3, the Arrhenius plot allows us to distinguish two different activation energies. A value of $E_a$ = 1.52 eV is obtained for intermediate temperatures, 600ºC < T < 800ºC, lower than the typical activation energies observed for pure LSM [28,42] and close to those obtained in LSM-YSZ composite cathodes [26,43,44]. For high temperatures, T > 800ºC, the activation energy reduces to almost half of the intermediate-temperature value, with $E_a$ = 0.86 eV. Therefore, our cathode exhibits two different operating regimes. Remarkably, the change of regime is at around 800ºC, which is very close to the temperature above which the Warburg characteristic slope is progressively lost. To sum up, at intermediate temperatures, the cathodic performance is dominated by a Warburg contribution with an activation energy close to the one obtained in cathodes with significant oxide-ion conduction while at higher temperatures, the activation energy reduces significantly.

A similar "two-regime" behaviour was observed in ceramic nanostructured LSM samples using Temperature Programmed Desorption and Thermogravimetric Analysis [35]. In that work, the authors concluded that the high activation energy at intermediate temperatures relates to a hindered oxygen diffusion through LSM, while on heating, a progressive release of oxide ions favours a fast oxygen transport.

Going back to Fig. 2, it can be seen that the imaginary part of the impedance drops very fast for frequencies, f < 100 Hz, and it is negligible for f < 10 Hz, even at relatively low temperatures. Also, a single impedance arc is observed in the whole temperature range. This feature is also found in composite LSM-YSZ cathodes [45] or MIEC materials [46]. Remarkably, we observe this effect in pure LSM cathodes, which usually show a significant low frequency contribution [23,26,27,46-50] generally attributed to gas phase transport [46]. In our case, the nanostructure seems to optimize the gas access to the surface making its contribution negligible. To confirm this, we performed EIS measurements at high and low $O_2$ partial pressures ($p(O_2)$) at 900ºC as shown in Fig. 4. We observe that the reduction of the oxygen partial pressure, induces the relative growth of the aforementioned low frequency process, evidenced as an additional arc in the Nyquist plot.

In order to separate the different contributions to the impedance according to their characteristic times, the impedance spectra can be adjusted using an equivalent circuit with components arising from contributions at different time scales.

At intermediate frequencies, the dominant process of the EIS spectra can be fitted using a finite length Warburg element (Ws) [41]. This element gives a contribution with a characteristic time that is related with a solid state diffusive process.

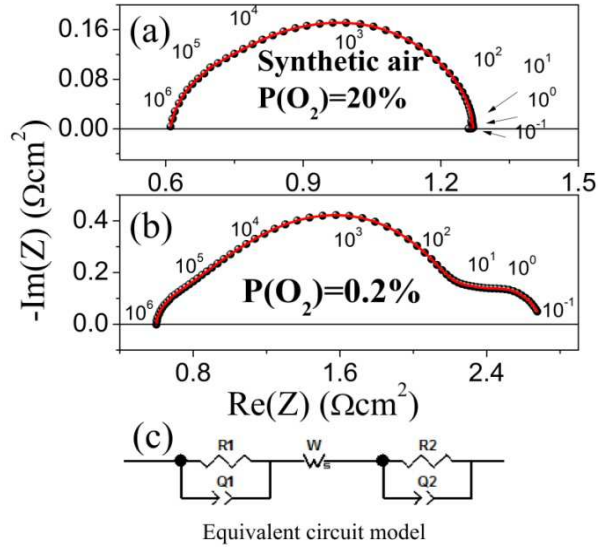

**Figure 4:** (color online) (a)-(b) Nyquist plots for different oxygen partial pressures p(O$_2$). Fittings with the model are presented in solid red line. (c) Schema of the equivalent circuit model.

Two additional processes should be considered to understand the EIS data, one at high frequencies and another one at low frequencies, which becomes evident at reduced oxygen pressure conditions (see Fig. 4). It is useful to consider them as arising from two components in series [41], each of them composed by the parallel of a resistance (R) and a constant phase element (Q). Each RQ component will account for a process with a characteristic time, in a similar way an RC circuit does. The constant phase element can be viewed as an imperfect capacitor, with a phase shift between 0º and 90º (0º will be a resistor and 90º an ideal capacitor) [41].

Thus, in order to account for the whole cathodic behavior, we used the equivalent circuit shown in Fig. 4 (c). This circuit has a small $R_1Q_1$ at high frequencies (~1MHz - 100kHz), most likely related to charge transfer processes between the electrode and the electrolyte [26,29,51], both in series with Ws and with an extra $R_2Q_2$ at low frequencies (~1-10 Hz). In Fig. 4 we show the fittings using the proposed circuit superimposed to the EIS spectra. Both, the data at atmospheric pressure (Fig. 4 (a)) and at reduced p(O$_2$) (Fig. 4 (b)) can be well adjusted by the proposed model.

In Fig. 5, we plot the resistive part of each process as a function of oxygen partial pressure at 900ºC. By comparing the dependence of the overall $R_p$ and the resistive part of the Warburg element ($R_W$), it is clear that the cathodic behavior is dominated by the Warburg process, according to their close values.

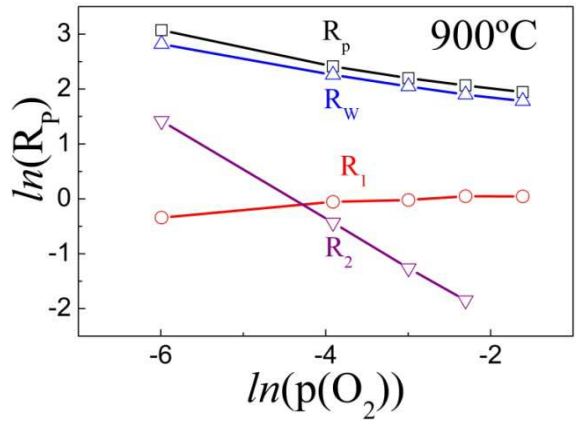

**Figure 5:** (color online) Oxygen partial pressure dependence of the resistive part of each of the circuit components of the impedance. (square) overall $R_p$, (circle) high frequency $R_1$, (up triangle) Warburg element $R_W$, (down triangle) low frequency $R_2$.

The resistive part of each component can be fitted using the classical power law dependence:

$R_i \alpha\ p(O_2)^{-n}$ (2)

where the subscript "$i$" stands for each process, while the value of "$n$" gives information about the type of species involved in the reaction that gives rise to this component [50].

The total polarization resistance can be well fitted with $n = 0.25 \pm 0.02$, which is a signature of the redox reaction between the adsorbed oxygen and the bulk electrode material [51,52].

The high frequency component ($R_1$) is nearly independent on $p(O_2)$, as expected for the charge transfer process between the electrode and the electrolyte [51]. The intermediate component ($R_W$) is the one that dominates the overall $p(O_2)$ dependence, as shown earlier. We will now focus on the low-frequency component ($R_2$) that presents the most significant change with $p(O_2)$ (see Fig. 5). As shown in Fig. 4, this component is negligible at atmospheric pressure, and it only becomes significant on reduced oxygen pressure conditions. The fitting analysis with Equation (2) gives a value of $n = 0.89 \pm 0.01$, indicating that it corresponds to gas phase diffusion [26,27,29,47,51,53]. Our results thus confirm that the nanostructure produces a qualitative change of the cathodic properties of LSM, making the contribution of gas phase transport negligible.

Our results can be summarized as follows: At intermediate temperatures (600ºC < T < 800ºC), impedance diagrams are similar to those for LSM-YSZ or MIECs, strongly suggesting that diffusion through the cathode is the limiting process in nanostructured LSM for this temperature range.

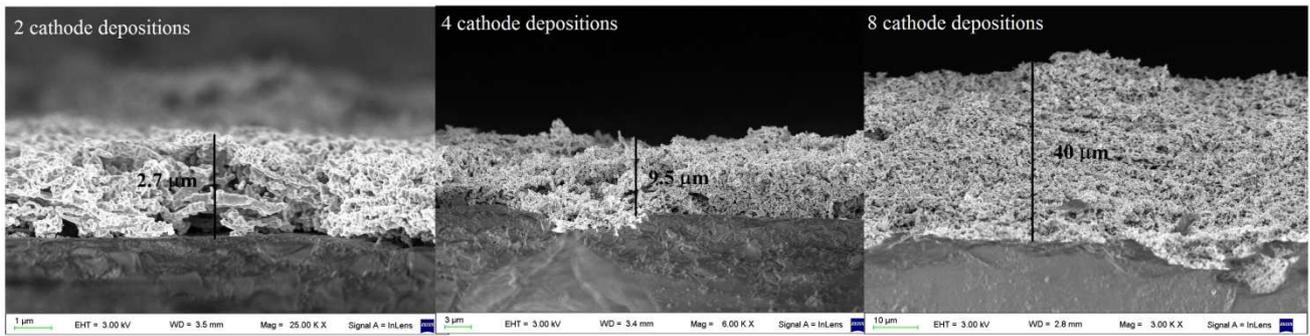

**Figure 6:** SEM images of the cross section of cathodes with different thickness from 2 to 8 cathode depositions.

At high temperatures (T > 800°C), the p($O_2$) dependence of the overall $R_p$ is consistent with a dominant redox reaction between the adsorbed oxygen and the electrode material. This means that, in this temperature regime, oxide ion transport through the cathode is so fast that the redox reaction at the surface becomes the rate-limiting step. The transition between the two regimes is evidenced in Fig. 3 by the change in activation energy from 1.52 eV to 0.86 eV, with the diffusion process dominating at intermediate temperatures, and the redox process being predominant in the high temperature range.

The role of the diffusive process through the cathode can be further confirmed by EIS measurements using cathodes with different thicknesses. We prepared samples performing 2 to 10 cathode depositions as described in the experimental section. The incremental depositions result in a thicker cathode, although some dispersion is observed on the thickness values, as can be seen in Fig. 6.

In Fig. 7 we show the polarization resistance as a function of the number of cathodes depositions. Our results show a reduction of $R_p$ when increasing the thickness of the cathode from 2 to 8 depositions, which then reverts when further increasing to 10 depositions. This result indicates that the cathode can not only behave as a pure electronic conductor. If that were the case, only the oxygen that was reduced in the TPB would reach the electrolyte. Given that the increase of the amount of cathode material obstructs the access of oxygen to the TPB, an improvement by adding more material unambiguously show that the oxygen reaching the electrolyte can be reduced in more remote areas than the TPB. This fact can only occur in the presence of diffusion of some oxygen species through the cathode. The increase in the resistance when reaching 10 depositions, could indicate the existence of a limit thickness beyond which the access of oxygen is so hard that even conduction through the cathode fails to compensate for the loss. However, further work is needed to elucidate this point.

Nanostructuring thus results in two improvements of the cathodic properties of LSM: it serves to overcome the limitation of gas phase transport of a porous LSM cathode and it enhances the oxide ion diffusivity through the cathode. Both effects contribute to an enhancement of the global ORR process. Our results also show that the redox reaction between the adsorbed oxygen and the electrode material, dominates the cathodic properties at high temperatures, due to the fast oxide ion transport.

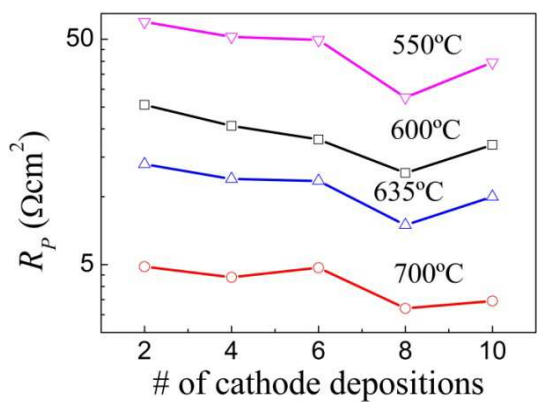

**Figure 7:** (color online) Polarization resistance at different temperatures as a function of the number of layers composing the cathode.

Having established that the cathodic properties of LSM involve significant oxygen diffusion through the cathode (not in the gas phase), it remains to understand the mechanisms underlying this particular behavior. The ORR process critically depends on the oxygen vacancy concentration, which according to previous works could be significantly enhanced at the surface [54,55]. To explore this point, we performed ab initio calculations for both, a slab and a bulk of $La_{0.8}Sr_{0.2}MnO_3$ studying the stability of oxygen vacancies close to the surface. Note that, due to explicit substitutional doping, our slab model is in fact asymmetric resulting in one surface of the slab having a slightly larger Sr content (denoted as Sr-rich) with respect to the other one (noted as Sr-poor). This allows us to study the influence of the Sr content at the surface in the vacancy formation energy ($E_{for}$) plotted in Fig. 8 as a function of the different planes of the slab. Taking into account the reference value for the bulk structure, we found, in agreement with previous theoretical studies for other perovskites [56], that the formation energy increases gradually from the surface towards the bulk-like layers, with an energy difference of up to 1.4 eV ($\Delta E_{surface}$) in favor of the formation of oxygen vacancies at the surface compared to bulk vacancies.

Furthermore, our simulations show that a Sr-rich environment decreases the formation energy and hence the Sr segregation would seem to favour the stabilization of oxygen vacancies at the surface [39]. The dispersion of formation energies obtained at different sites within each $MnO_2$ layer could also be attributed to different Sr environments. These issues foster further investigation both from the theory and experimental viewpoints.

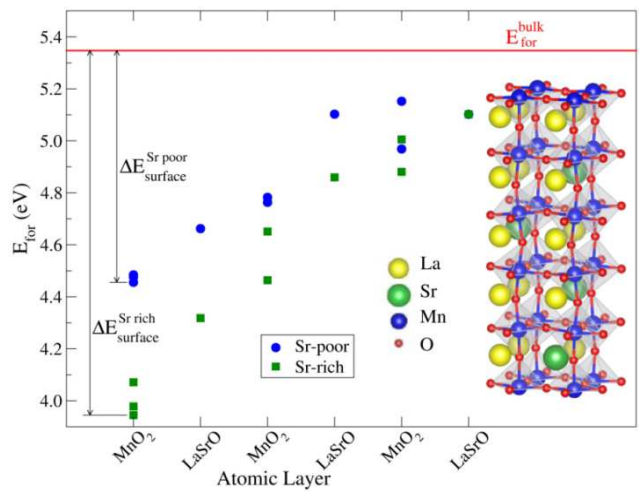

**Figure 8:** (color online) Formation energy ($E_{for}$) for oxygen vacancies within a $La_{0.8}Sr_{0.2}MnO_3$ slab. The unit cell used in the calculation is included. The horizontal axis denotes the atomic layers of the slab, starting from the surface (to the left) towards the bulk-like planes (to the right). The red solid line indicates the reference value for the vacancy formation energy obtained in a bulk calculation. The green(blue) squares (circles) indicate the Sr-rich (Sr-poor) side of the slab. $\Delta E_{surface}$ highlights the energy gain at the surface in each case.

**Conclusions**

Summarizing, we have studied the electrochemical properties of nanostructured LSM cathodes, finding that their enhanced performance (in comparison to microstructured cathodes) is promising for their use as IT-SOFCs. Cathodic properties are dominated by ionic conduction instead of gas phase transport as a consequence of using a porous nanostructure. The presence of ionic conductivity in nanostructured LSM was confirmed by EIS experiments as a function of thickness. Our cathodes operate in two distinct regimes (both of them exhibiting negligible oxygen gas transport contribution): at intermediate temperatures (up to 800ºC), the electrochemical properties are dominated by ionic conduction while, at high temperatures, the limiting process is the redox reaction between the adsorbed oxygen and the bulk electrode material.

Although the appearance of oxide ion conduction was not previously reported in LSM SOFC cathodes, our study is inspired in previous findings for LSM. Hammouche et al. [57] ascribed the large electrocatalytic properties of LSM to the generation of oxygen vacancies that are mobile enough to carry oxide ions from the surface of the electrode to the electrolyte. Also, evidences of ionic transport through dense LSM cathodes have been shown in earlier studies [58,59]. In addition, fast diffusion of oxygen at the grain boundaries has been observed by the isotope exchange depth profiling technique using SIMS depth profiling in microcrystalline A-site deficient La manganites [32] and in LSM [33]. Thus, in our case, the disorder at the surface of the nanoparticles that compose the cathode is probably causing an extension of the ORR active zone via the formation of regions with a high concentration of oxygen vacancies (as evidenced by ab initio simulations), giving rise to significant ionic conduction. The large surface-to-volume ratio in our cathode is most likely to be responsible for that behaviour.

To the best of our knowledge, this is the first time that the performance of nanostructured LSM cathodes is reported. More interestingly, our EIS study for different temperatures and oxygen partial pressures allowed the elucidation of the mechanisms involved in the ORR and clearly demonstrated that significant oxide ion conductivity arises due to nanostructuration. This remarkable qualitative change in the conduction mechanism, from electronic conduction in microcrystalline LSM to mixed electronic-ionic conduction in nanocrystalline LSM, should inspire future works on other cathodes, which could also improve their electrochemical performance via nanofabrication techniques.

**Acknowledgements**

Financial support from CONICET (PIP00038) and ANPCyT (PICT 1327, 1948 and 2689) is acknowledged. J.H.S. is currently at TENARIS, Siderca (Argentina). A.R.L. is currently at FCEN-UBA (Argentina). MP acknowledges support from MINECO (FIS2015-64886-C5-3-P and SEV-2013-0295) and Generalitat de Catalunya (2014SGR301). We thank Paula Abdala for her assistance with samples preparation and EIS measurements, Solange Di Napoli and Cecilia Fischer for manuscript revision and Lucía Sacanell Rio for her assistance with the graphical abstract.

**Graphical Abstract:**

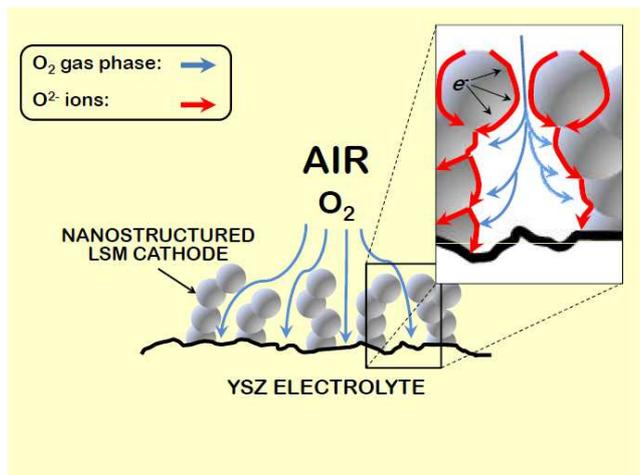